\documentclass[fleqn,10pt]{wlscirep}
\usepackage[utf8]{inputenc}
\usepackage[T1]{fontenc}

\usepackage[nomarkers,nofiglist,notablist]{endfloat}

\title{Spatiotemporal shaping of broadband helical light pulses at relativistic intensities}

\author[1,*]{Andrew Longman}
\author[1,2]{Danny Attiyah}
\author[1]{Elizabeth S. Grace}
\author[3]{Christopher Gardner}
\author[1]{Tayyab Suratwala}
\author[1]{Gary Tham}
\author[1]{Colin Harthcock}
\author[4]{Robert Fedosejevs}
\author[2]{Franklin Dollar}

\affil[1]{Lawrence Livermore National Laboratory, Livermore, 94550, USA}
\affil[2]{STROBE, NSF Science \& Technology Center, University of California, Irvine, 92697, USA}
\affil[3]{University of California Irvine, Department of Physics and Astronomy, Irvine, 92697, USA}
\affil[4]{University of Alberta, Department of Electrical and Computer Engineering, Edmonton, T6G1R1, Canada}

\affil[*]{longman1@llnl.gov}

\begin{abstract}
Spatiotemporal control of laser pulses at relativistic intensities is a longstanding goal with broad implications in laser-plasma acceleration, high-brightness radiation sources, and extreme-field science. Laser pulses with helical spatiotemporal intensity profiles, often referred to as light springs, carry multiple spectral and orbital angular momentum (OAM) modes, producing a rotating intensity profile capable of coupling directly to helical plasma waves. Until now, light springs have only been realized on low-power systems, limited by optical damage thresholds and large-aperture beamline constraints. Here, we report the first experimental realization of light springs at relativistic intensities, achieving peak intensities above $1.4\times10^{18}$ W/cm$^{2}$. Our approach spectrally splits a high-power laser pulse, imprints distinct helical phases on each component, and coherently recombines them. Hyperspectral imaging, off-axis holography, and spectral phase reconstruction confirm excellent agreement with theory and reveal the potential to drive superluminal rotational velocities. Introducing spectral chirp demonstrates further control of the temporal evolution of the transverse mode structure. This platform opens new regimes of ultra-intense laser–plasma interaction where laser OAM can directly couple to plasma OAM.

\end{abstract}
\begin{document}
\flushbottom
\maketitle

\thispagestyle{empty}

\section*{Introduction}
The spatiotemporal shaping of ultrafast (<100 fs) light pulses has emerged as a powerful tool for generating, probing, and manipulating matter \cite{Forbes2021}. By combining spectral and spatial phase-shaping techniques, the generation of laser pulses with arbitrary intensity and phase profiles in both space and time has now become possible \cite{Li:18,Bliokh_2023,Shen_2023,Jhajj16,Hancock:19}. Of particular interest are pulses whose transverse intensity envelope gyrates about the propagation axis, a-so-called spatiotemporal "light spring". Such fields arise from the interference of two co-propagating, helically phased beams of different colors \cite{Pariente:15,Zhao2020,Cruz-Delgado2022,Chen22,Piccardo2023,Oliveira25,Mendonca22}. Light springs possess unique properties, including programmable transverse velocity, time-dependent orbital angular momentum (OAM), and when intense, have the ability to drive helical plasma waves \cite{Blackman2019,Blackman2020,Vieira14,Palastro24}. In contrast, laser pulses carrying a single OAM mode exhibit azimuthally symmetric intensity profiles, yielding a ponderomotive force with no azimuthal component and, to first order, no ability to drive helical plasma waves. Theoretical studies suggest that such helical plasma waves can enhance laser–plasma interactions by enabling rotating wakefields \cite{Vieira2018,Mendonca22}, novel radiation sources \cite{Rego19,Mendonca24}, and the generation of strong axial magnetic fields \cite{Shi2018}. These applications generally require relativistic intensities with normalized vector potentials on the order of $eE_0/m_ec\omega_0=a_0\sim1$, where $e,E_0, m_e, c, \omega_0$ are the electron charge, electric field of the laser, electron mass, speed of light, and laser frequency, respectively.

While prior work has demonstrated spatial or temporal control of relativistic pulses, achieving full spatiotemporal control remains challenging. Spectral phase modulation via programmable dispersive filters offers in-situ control of temporal dispersion and pulse-shaping \cite{TOURNOIS1997}, while "flying-focus" techniques \cite{Sainte-Marie:17,Froula2018,Liberman:24} can control the velocity and direction of a pulse’s intensity envelope. Laguerre-Gaussian (LG) mode conversion at high-power has been made possible using recent developments in off-axis spiral phase mirrors (SPMs) and spiral phase plates, enabling the first experimental studies of relativistic-intensity single-mode OAM beams \cite{Brabetz2015,Longman:20a,Wang2020,Lee:24,Longman2022,Longman:20b}. 

To date, spatiotemporal light springs have been generated by spatially dispersing the beam, typically using gratings or axicons, and imprinting OAM onto individual spectral components using spatial light modulators (SLMs), phase masks, metamaterials, or high harmonic generation \cite{Cruz-Delgado2022,Piccardo2023,Chen22,Divitt19,Hernandez13}. However, these methods are unsuitable for high-power systems due to the limited size and damage thresholds of SLMs and metamaterials, as well as the large beam diameters involved, which often exceed 150 mm in the near field. Generating relativistic-intensity light springs requires near-field optics that can withstand ultrashort pulse fluences on the order of 0.1 J/cm$^2$, exhibit minimal dispersion, and scale to large-aperture, high-energy laser systems without requiring additional vacuum chambers or large diffraction gratings.

In this work, we present the first experimental platform capable of generating and characterizing spatiotemporal light springs at relativistic intensities ($a_0\sim1$).  These pulses exceed previous intensity benchmarks by several orders of magnitude and are, in principle, scalable to the large apertures required for petawatt-class laser systems. We demonstrate three distinct broadband light springs with $a_0\approx1$, in which we control the handedness of the light spring, its modal evolution in time, and its transverse velocity, including the possibility of driving superluminal light springs. These high-intensity spatiotemporal structures provide a versatile tool for driving structured plasma waves and open new avenues for investigating OAM-mediated laser–plasma interactions.

\section*{Results}
\begin{figure}[h]
\centering
\includegraphics[width=\linewidth]{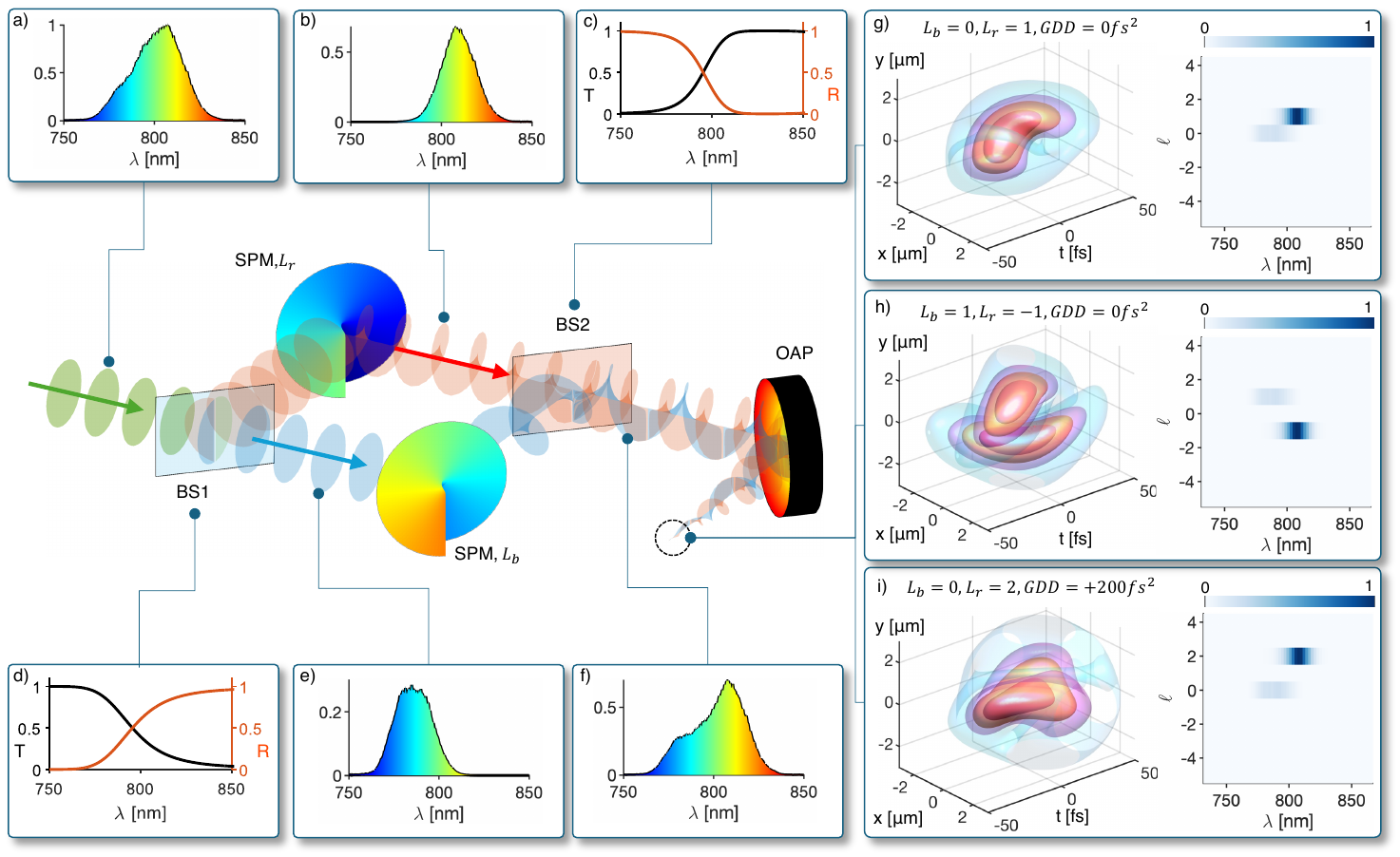}
\caption{Generation of high-intensity broadband light springs. A compressed broadband laser pulse (a) is spectrally split into red (b) and blue (e) components using a short-pass dichroic beamsplitter (BS1, d). Each respective beam is the reflected off an off-axis spiral phase mirror (SPM), denoted as $L_r$ and $L_b$. The two beams are recombined using a long-pass dichroic beamsplitter (BS2, c) and subsequently focused by an off-axis parabola (OAP), producing a spatiotemporal structured focal spot. Subplots (g-i) show intensity isosurfaces (0.5, 0.25, 0.125, and 0.025 of peak) and spectral-topological correlations for three configurations: $L_b=0, L_r=1, GDD=0$ fs$^2$, $L_b=1, L_r=-1, GDD=0$ fs$^2$, and $L_b=0, L_r=2, GDD=+200$ fs$^2$.}
\label{fig1}
\end{figure}

Our proof-of-principle platform manipulates a broadband laser pulse in the near-field, specifically after amplification and compression but prior to the focusing optics, using four primary optical components, as illustrated in Fig. \ref{fig1}. The beam is first split by a short-pass dichroic beam-splitter (BS1, Fig.\ref{fig1}d) whose transmission and reflection inflection points are centered near the laser's center wavelength $\lambda_0$. To minimize dispersion and nonlinear phase accumulation, BS1 and BS2 are fabricated from 1 mm thick fused silica polished to $\lambda/10$ flatness. These introduce approximately 50 fs$^2$ of group delay dispersion (GDD) and a B-integral of $\sim$0.3 at 45$^\circ$ incidence. The spectral transition regions of the coatings are engineered to limit second- and third-order dispersion to less than or equal to 200 fs$^2$ and 2000 fs$^3$, respectively, ensuring that the pulse duration remains below 50 fs. Additional details on the beamsplitter designs are provided in the Methods section. 

The two beams, referred to as "red" and "blue", are directed onto separate off-axis spiral phase mirrors (SPMs), labeled $L_r$ and $L_b$, respectively \cite{Longman:20a}. These mirrors impart a helical phase to the beams, thereby introducing orbital angular momentum (OAM), with a topological charge given by $L=2H/\lambda_0\cos(\theta_i)$ where $H$ is the step height, $\lambda_0$ is the central laser wavelength, and $\theta_i$ is the incidence angle \cite{Longman:20a,Longman2022}. By designing $H$ such that $L$ is an integer, the SPM efficiently converts the incident Gaussian beam into a dominant Laguerre-Gaussian (LG) mode with azimuthal index $\ell=L$. Owing to the beams' broad spectral bandwidth and their spectral separation, the resulting fields form a superposition of low-order radial modes ($p$) along with dominant azimuthal modes, characterized by $\ell_r=L_r,\ell_b=L_b$ for the red and blue arms, respectively \cite{Longman:17}. 

The beams are recombined by a long-pass dichroic beamsplitter (BS2, Fig.\ref{fig1}c), which is designed with dispersion characteristics matched to those of BS1. Although the recombination process introduces some optical loss, primarily due to spectral clipping at the overlap region, careful optimization of the dielectric coatings yields an overall system efficiency of approximately 70$\%$, with potential for further improvement. The combined output spectrum is shown in Fig.\ref{fig1}f, normalized to the integrated input spectrum displayed in Fig.\ref{fig1}a. The superimposed beams are subsequently directed to an off-axis parabolic mirror (OAP), which focuses them to a high-intensity spatiotemporal light spring. 

Given the amplitude and phase of each spectral component in the near-field, the corresponding focal spot in the far-field can be reconstructed by summing the spatial Fourier transforms of the complex near-field. Further details are provided in the Methods section. Simulated intensity isosurfaces for two combinations of topological charge, as well as one case with chirped spectral phase, are shown in Figs.\ref{fig1}g-\ref{fig1}i. These simulations incorporate the experimental laser spectrum along with the measured transmission and reflection curves of the beamsplitters. In each plot, the isosurfaces correspond to the relative intensity levels of 0.5, 0.25, 0.125 and 0.025 of the peak intensity. 

Figure \ref{fig1}g shows a light spring formed with parameters $L_r=1,L_b=0$, and zero GDD. The resulting spatiotemporal structure takes the form of a single-fold helix twisting about a central axis. Videos showing the transverse focal plane intensity as a function of time are provided in the Supplementary Material for all light springs shown in this work. To the right of the intensity isosurface, the spectral-topological correlation plot illustrates the relative energy in each $\ell$ mode as a function of wavelength. As expected, the energy is concentrated primarily in the $\ell=0$ and $\ell=1$ modes. In this configuration, the beam splitters were designed with central wavelengths near 795 nm, resulting in greater energy in the red portion of the spectrum. 

Figures \ref{fig1}h and \ref{fig1}i show intensity isosurfaces for light springs formed with $L_r=-1,L_b=1$ and $L_r=2,L_b=0$, respectively. The number of intensity lobes in the transverse plane corresponds to the absolute difference in topological charge, $\Delta\ell=\ell_b-\ell_r$. Accordingly, Fig.\ref{fig1}h) exhibits two rotating lobes. Notably, the direction of rotation is opposite to that of the light spring in Fig.\ref{fig1}g, due to the reversed slope of the spectral-topological correlation, $d\ell/d\lambda $. In Fig.\ref{fig1}i, the beam is chirped with a GDD of +200 fs$^2$, resulting in a pulse with a time-varying OAM that evolves from an $\ell=0$ mode to an $\ell=2$ mode.

To demonstrate relativistic strength light springs in the laboratory, we used a 15 mm diameter, 7 mJ, 34 fs Ti:Sapphire laser pulse with a quasi-Gaussian spectrum centered at 800 nm (Fig.\ref{fig1}a). The transmission and reflection curves of the dichroic beamsplitters are shown in Fig.\ref{fig1}c and \ref{fig1}d for the long-pass and short-pass elements, respectively. A spiral phase mirror with topological charge $L=1$ (Fig.\ref{fig3}a), was placed in each arm alternately with a flat mirror, enabling two configurations: $L_r=1,L_b=0$, and $L_r=0,L_b=1$. The beams were focused using an OAP with $f_{\#}=2$, and characterized using a combination of a frequency resolved optical gating (FROG) to measure the spectral phase and intensity, along with hyperspectral imaging and off-axis holography to measure the spatial phase as a function of wavelength. Further details of this measurement process is provided in the Methods section.  

\begin{figure}[h]
\centering
\includegraphics[width=\textwidth/2]{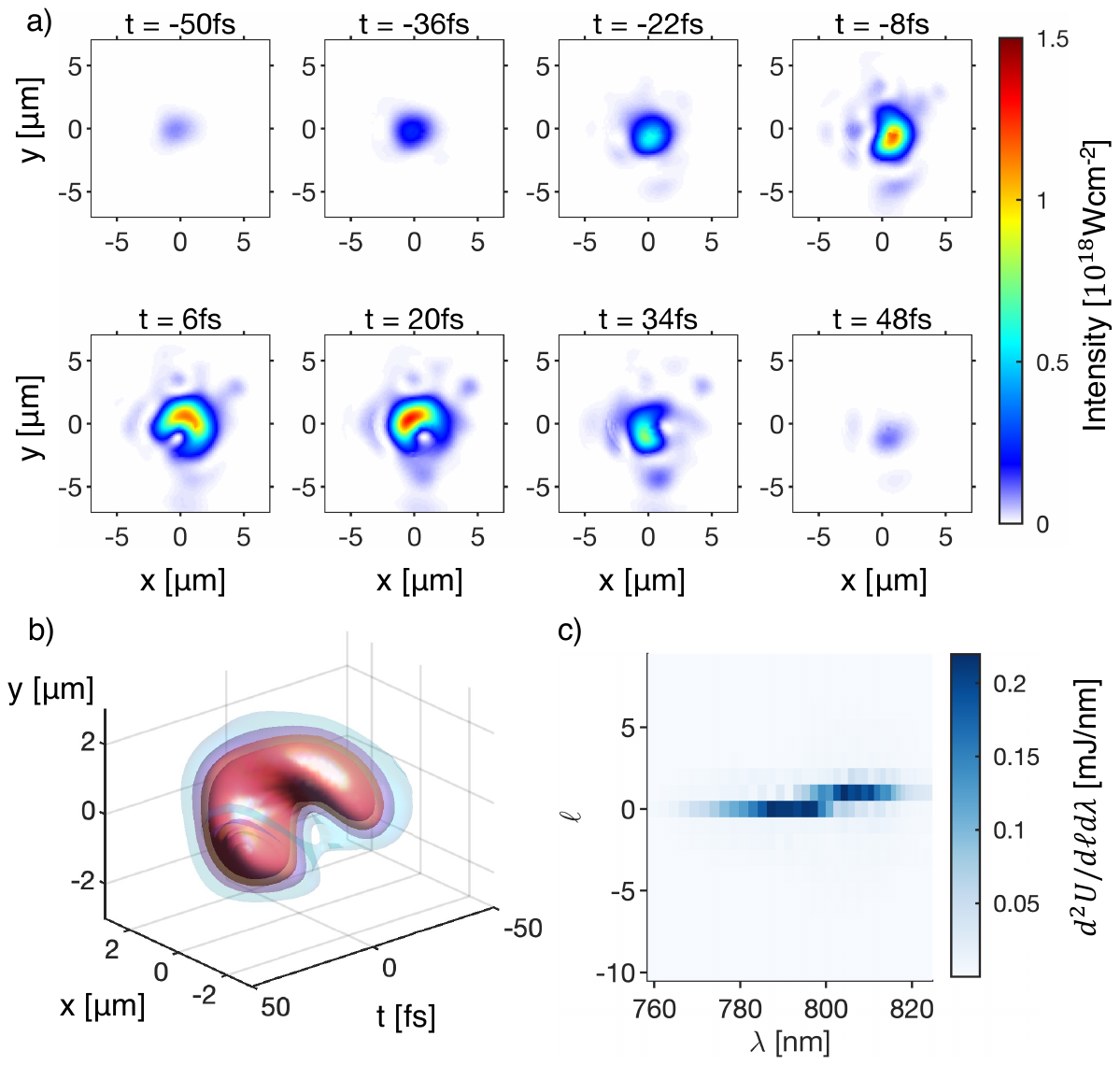}
\caption{Measurements of a light spring with $L_r=1,L_b=0$ with GDD = 120 fs$^2$. (a) Temporal evolution of the intensity in the focal plane, shown in units of $10^{18}$ Wcm$^{-2}$. (b) Intensity isosurface plot with surfaces corresponding to 0.5, 0.375, 0.25 and 0.125 of the peak intensity. c). Spectral-topological correlation, with each mode normalized to its absolute energy.}
\label{fig2a}
\end{figure}

Utilizing the measurements from the FROG and the hyperspectral imaging system, we are able to reconstruct the electric field of the laser pulse as a function of time and space. The results for the $L_r=1,L_b=0$ configuration are given in Fig.\ref{fig2a}. Eight time slices of the laser pulse intensity are displayed in Fig.\ref{fig2a}a, with the estimated intensity indicated by the color bar. The peak intensity exceeds $1.4\times10^{18}$ W/cm$^{2}$, corresponding to a normalized vector potential of $a_0\approx0.8$. The corresponding isosurface plot in Fig.\ref{fig2a}b clearly reveals the spatiotemporal helical structure characteristic of a light spring.  The spectral-topological correlation (Fig.\ref{fig2a}c) shows distinct spectral separation between two dominant OAM modes with a positive slope $d\ell/d\lambda$. To minimize the pulse duration, the compressor gratings were adjusted to compensate the for beamsplitter dispersion, resulting in a residual GDD of 120 fs$^2$, corresponding to a pulse duration of 47 fs (FWHM).

\begin{figure}[h]
\centering
\includegraphics[width=\textwidth/2]{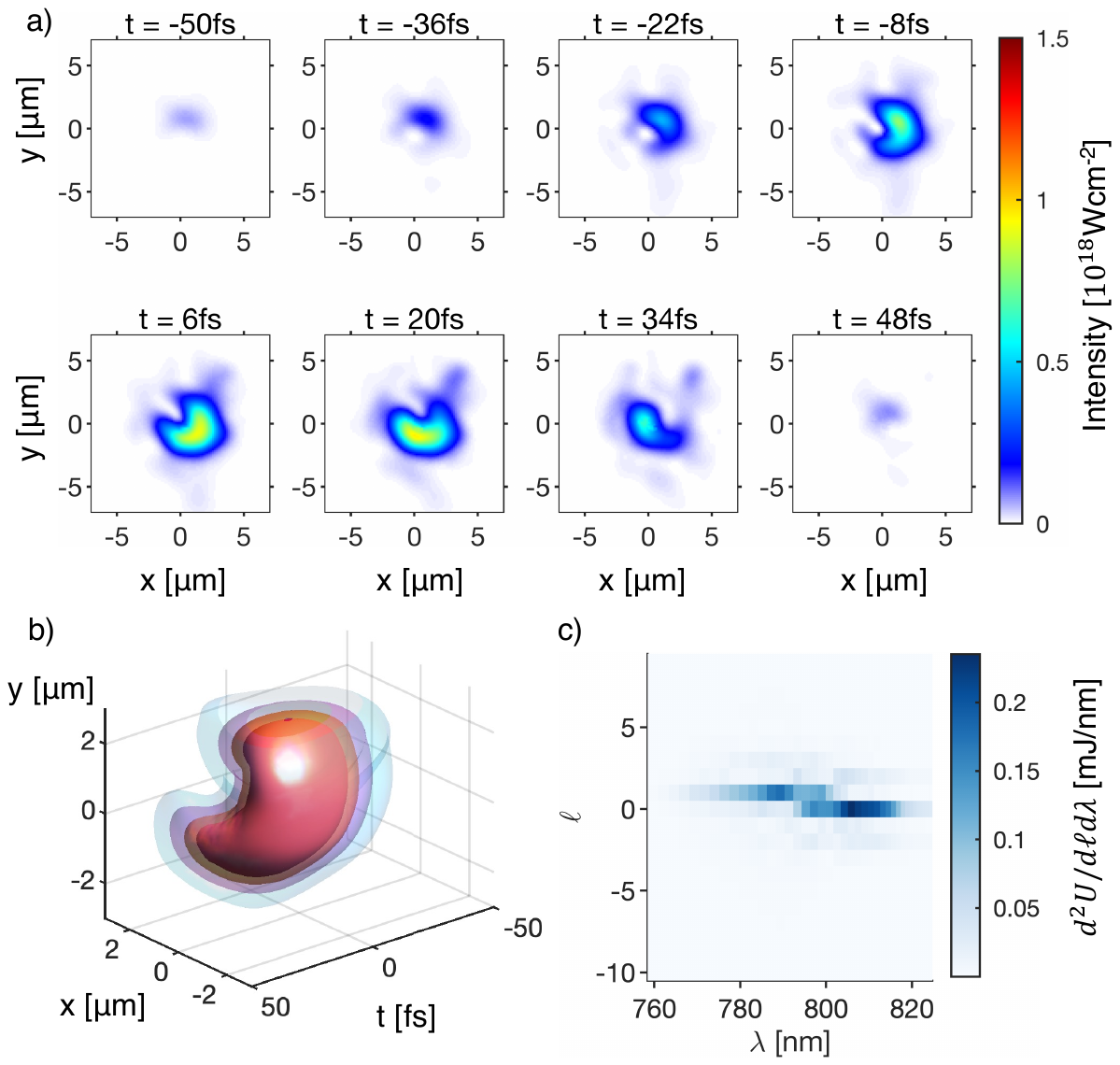}
\caption{Measurements of a light spring with $L_r=0,L_b=1$ with GDD = 120 fs$^2$. (a) Temporal evolution of the focal plane intensity, shown in units of $10^{18}$ W/cm$^{2}$. (b) Intensity isosurfaces at 0.5, 0.375, 0.25 and 0.125 of the peak. c). Spectral-topological correlation, normalized to its absolute energy.}
\label{fig2b}
\end{figure}

As shown in the eight time slices of the pulse intensity (Fig.\ref{fig2a}a), the modes at each time slice, and equivalently in spectral space, exhibit complex structures. To calculate the angular momentum as a function of wavelength, precise knowledge of the beam axis is essential; otherwise, a significant offset in the measured OAM, denoted $\ell_{\mathrm{ext}}$, can arise due to extrinsic contributions. Given the asymmetric nature of the focal spot intensity, we determined the beam axis by locating the centers of the pure modes at the spectral extrema, where the beam axis is well defined, and then interpolated this axis across the intermediate spectral regions where significant mode mixing occurs. Further details regarding the extrinsic and intrinsic contributions to the spectral-topological correlations are given in the Methods section.

Interchanging the spiral phase mirrors to generate a light spring with $L_r=0,L_b=1$ reverses the helicity, as evidenced by the negative slope $d\ell/d\lambda$ in Fig.\ref{fig2b}c. In this configuration, the peak intensity is slightly reduced relative to Fig.\ref{fig2a}, primarily due to increased energy in the $\ell=0$ mode. This arises from the beamsplitter central wavelengths being 795 nm, leading to asymmetric energy partitioning. The resulting peak intensity is approximately $0.9\times10^{18}$ W/cm$^{2}$, corresponding to $a_0\approx0.6$.

\begin{figure}[h]
\centering
\includegraphics[width=\textwidth/2]{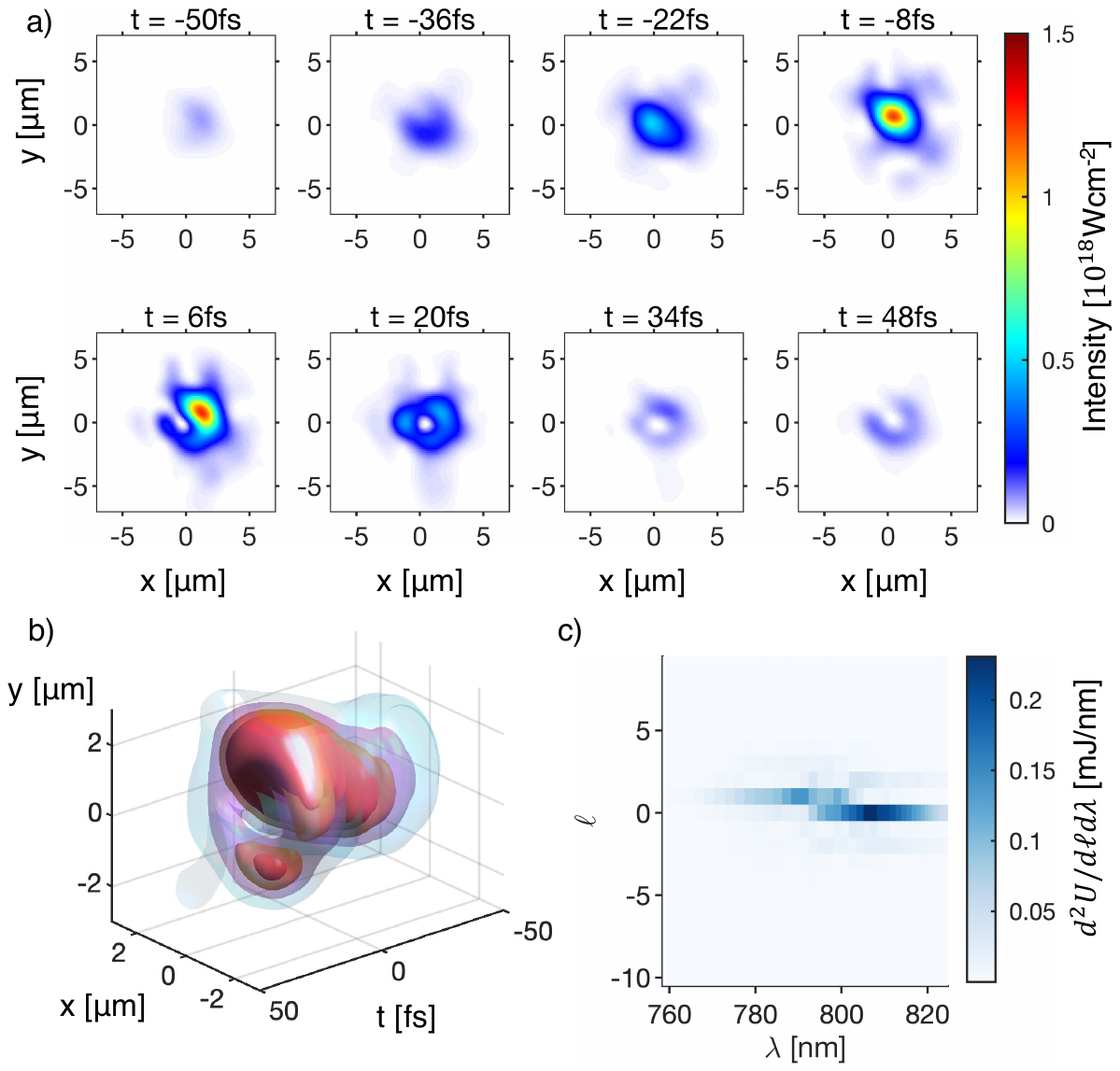}
\caption{Measurements of a negatively chirped light spring with $L_r=0,L_b=1$ and GDD = -1490 fs$^2$. (a) Temporal evolution of the intensity in the focal plane, shown in units of $10^{18}$ W/cm$^{2}$. (b) Intensity isosurface plot, with surfaces corresponding to 0.275, 0.21, 0.14 and 0.07 of the peak intensity. c). Spectral-topological correlation, with each mode normalized to its absolute energy.}
\label{fig2c}
\end{figure}

Shifting the central wavelength of the beamsplitter, $\lambda_b$, away from the laser’s central wavelength introduces a power imbalance between the two arms, as shown in Fig.\ref{fig1}. At first glance, this imbalance may appear undesirable. However, the resulting redistribution of energy can significantly modify the properties of the light springs, as demonstrated in the experimental results in Figs.\ref{fig2a} and \ref{fig2b}. This effect provides an additional degree of control and can be used to tailor the characteristics of the light springs.

Using this same mirror configuration, we intentionally introduced negative chirp to generate a light spring with time-varying OAM content, as shown in Fig.\ref{fig2c}. In this case, the pulse was negatively chirped such that the red component ($\ell=0$) arrives earlier that the blue component ($\ell=1$), resulting in a beam whose OAM mode evolves dynamically in time. Despite the added dispersion, the beam remains at relativistic intensity, with a peak intensity of approximately 1.3$\times10^{18}$ W/cm$^{2}$, comparable to that of the best-compressed light springs.  

As demonstrated, it is possible, in principle, to program light springs by controlling their spectral and spatial parameters. The number of rotating intensity lobes in the transverse plane is determined by the absolute topological charge difference, $\Delta\ell=\ell_b-\ell_r$, while the direction of rotation is set by the sign of $d\ell/d\lambda$. Additionally, the overall shape of the light spring can be tuned by adjusting the center wavelength of the beamsplitter. This degree of control enables precise sculpting of the spatiotemporal intensity pattern, which directly translates into dynamic ponderomotive forces with spatially and temporally varying orientations. The transverse rotation speed of these features can be approximated by dividing the focal spot circumference by the orbital period, yielding:
\begin{equation}
    \frac{v_\perp}{c}\approx 2\pi f_{\#}\frac{\Delta\lambda}{\lambda_0}
\end{equation}
Using a laser bandwidth of approximately 30 nm and an $f_{\#}$ of 2, we estimate a transverse rotational velocity of $v_{\perp}/c\sim0.5$. From Fig.\ref{fig2a}a, the light spring rotates by approximately $\pi/2$ rad between 6fs and 20fs. Estimating the radius from the figure to be approximately $1.5$ $\mu$m, this corresponds to a rotational velocity of $v_\perp/c\approx0.56$, in good agreement with the theoretical estimate. 

Importantly, increasing either the focusing parameter $f_{\#}$ or the laser bandwidth allows the transverse rotational velocity to exceed the speed of light, resulting in superluminal motion of the intensity pattern. While this does not violate the principles of relativity, since no physical energy or information propagates faster than light, such apparent superluminal motion enables a range of novel optical and plasma phenomena. For instance, flying-focus pulses can use ponderomotive potentials with superluminal velocity to support dephasingless electron acceleration \cite{Palastro20}, allowing particles to remain phase-locked to the intensity maxima over extended distances, thereby enhancing both energy gain and beam stability. Similar dephasingless schemes may be realizable using light springs to control the transverse velocity \cite{Vieira2018}. These conditions may also enable the excitation of new classes of nonlinear plasma waves and facilitate exotic particle trapping mechanisms that are inaccessible using conventional Gaussian and OAM beams.

Furthermore, the temporal evolution of the OAM mode can be precisely controlled by introducing spectral chirp through fine adjustments to the grating separation in the laser compressor. This approach provides an additional degree of freedom for engineering the spatiotemporal structure of the pulse, allowing the focal spot size to expand or contract dynamically in time. Such temporal zooming modifies the local intensity and ponderomotive force profile throughout the pulse duration, enabling new regimes of plasma control, such as manipulating ionization front velocities or modulating wakefield properties in real time. Additionally, this chirp-induced evolution alters the number and structure of focal spot lobes in the far field, influencing the angular momentum transfer to the plasma and allowing selective excitation of specific plasma modes or tailored injection conditions for electron beams.

\section*{Discussion}
The generation and spatiotemporal characterization of relativistic-intensity light springs presented in this work establishes a new regime of structured light, in which spatial and temporal degrees of freedom are intrinsically coupled through an engineered OAM spectrum. By extending the light spring concept into the relativistic domain and achieving normalized vector potentials exceeding $a_0 \approx 0.8$, we show that both the helicity and modal composition of the beam can be precisely tailored and resolved with high fidelity in both space and time.

This advance is enabled by the development of broadband spiral phase optics and custom-designed dichroic beam splitters, which provide fine control over the joint spectral and azimuthal content of the beam. A key strength of this approach is its scalability. Spiral phase mirrors with apertures exceeding tens of centimeters have already been demonstrated \cite{Wang2020,Longman:20a}, and current advances in polishing and coating technologies make it feasible to fabricate multi-region phase structures compatible with large-aperture, high power beamlines. 


The combination of relativistic intensity, tunable OAM content, and femtosecond-scale temporal structure enables a wide range of new applications. These include the generation of high harmonics with time-varying OAM, the use of the helical ponderomotive forces for coupling to helical plasma waves \cite{Shi2018}, and the excitation of twisted instabilities in laser-plasma interactions \cite{Blackman2019}. Looking ahead, the interaction of such beams with nonlinear optical media and dense plasmas offers a rich experimental platform for exploring the transport, coupling, and conservation of angular momentum under extreme field conditions.

\section*{Methods}
\begin{figure}[h]
\centering
\includegraphics[width=\linewidth]{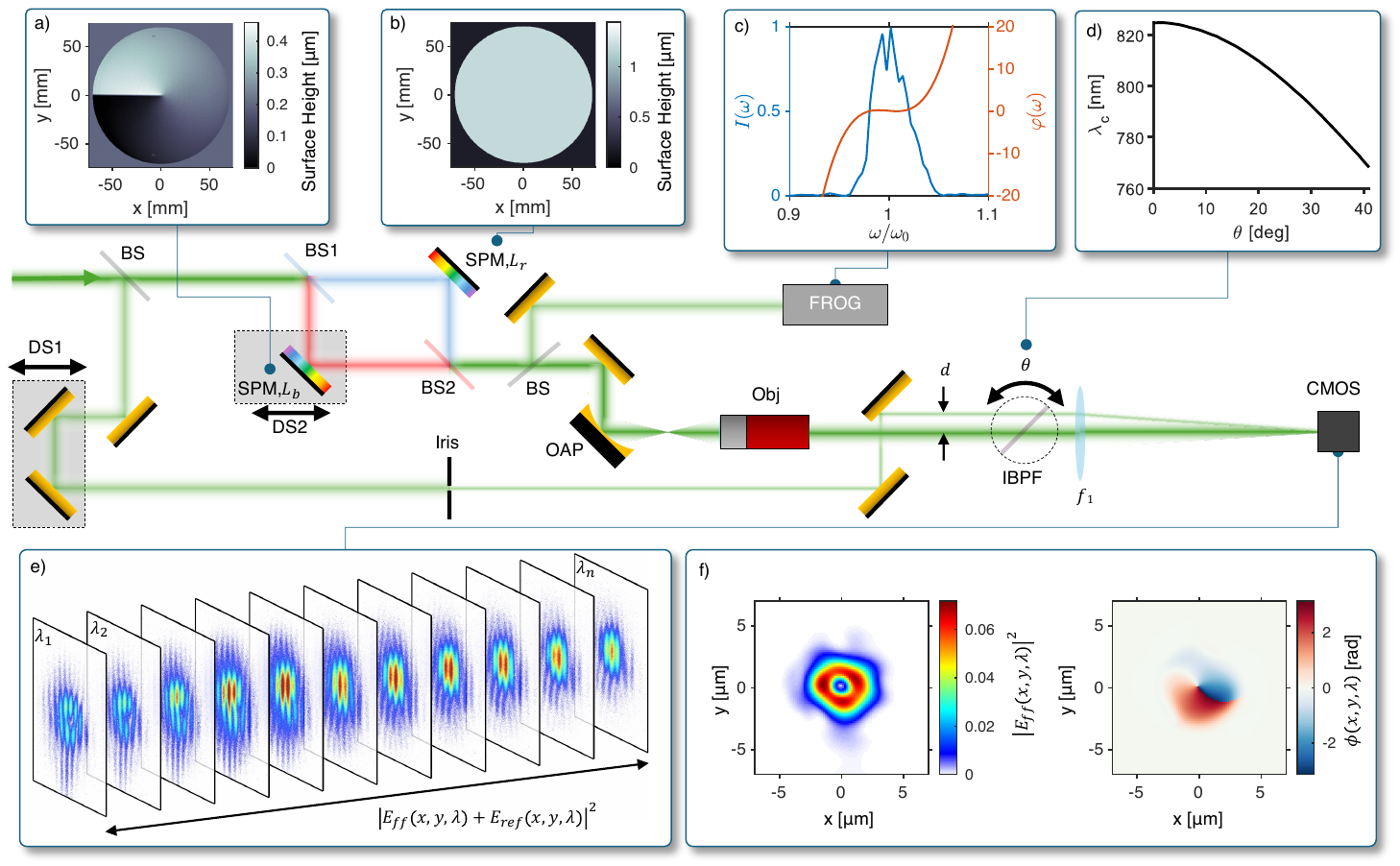}
\caption{Experimental platform for generating and measuring relativistic-intensity light springs. The beam is first split by an intensity beam splitter (BS), with the transmitted portion spectrally separated by a short-pass dichroic beam splitter (BS1). Each spectral component reflects off a spiral phase mirror (SPM), shown in (a) for $L=1$ and (b) for the $L=0$. The beams are then recombined using a long-pass dichroic beam splitter (BS2). A second BS picks off a portion for spectral phase analysis via frequency resolved optical gating (FROG) (c). The transmitted beam is imaged onto a high resolution CMOS sensor using a 20x long-working-distance objective and a $f_1=1$ m lens, with a tunable interference band-pass filter (IBPF) selecting the focal spot wavelength (d). A reference beam, delayed via a stage (DS1) and apodized to 0.5 mm, interferes with the main beam to generate holograms of each wavelength (e), enabling amplitude and phase extraction (f).}
\label{fig3}
\end{figure}

Measurement and characterization of the spatiotemporal light springs can generally be performed using various spatiotemporal field reconstruction methods \cite{Jolly_2020,Pariente2016}. In our experiment, we employed a multi-shot implementation of STRIPED FISH \cite{Guang:17}, which reconstructs the electric field in the focal-plane from three primary measurements: a frequency resolved optical gating (FROG) trace, and a combined hyperspectral imaging system with off-axis holography, as summarized in Fig.\ref{fig3}. 

In this scheme, the time-resolved electric field of the laser pulse is reconstructed by taking the inverse Fourier transform of the frequency-resolved electric field measured in the far-field,
\begin{equation} \label{eq1}
\tilde{E}_{LS}(x,y,\omega) = E_0(\omega)\, \psi_{ff}(x,y,\omega)\, \exp\left[-i\varphi(\omega)\right]
\end{equation}
where $E_0(\omega)$ and $\varphi(\omega)$ are the spectral amplitude and phase of the pulse, obtained from a FROG measurement, and $\psi_{ff}(x,y,\omega)$ is the complex spatial field in the far-field, measured via off-axis holography and hyperspectral imaging. 

The time-dependent electric field is then given by 
\begin{equation}
    E_{LS}(x,y,t)=\mathcal{F}^{-1}\left\{\tilde{E}_{LS}(x,y,\omega)\right\}.
\end{equation}
The Fourier transform imposes constraints on the spectral and temporal resolution through the standard relationships, $\Delta\omega=2\pi/T$ and $\Delta t=2\pi/\Omega$, where $T$ and $\Omega$ are the temporal and spectral windows respectively. For example, a 1 ps time window corresponds to a maximum spectral sampling interval of $\Delta\omega\leq6.2\times10^{12}$ Hz, or equivalently, $\Delta\lambda\approx2.1$ nm at 800 nm.

\subsection*{Hyperspectral imaging, off-axis holography, and spatiotemporal reconstruction}
To resolve the focal intensity distribution at each wavelength with $\Delta\lambda\leq2.1$ nm, we employed a narrowband interference bandpass filter (IBPF) with a nominal transmission window of 1 nm and center wavelength $\lambda_{CWL}=$825 nm. The IBPF was rotated about its optical axis to scan the transmitted central wavelength according to the relation 
\begin{equation}
    \lambda_{c}=\lambda_{CWL}\sqrt{1-\frac{\sin^2\theta_i}{n_{eff}^2}}
\end{equation}
where $n_{eff}\approx1.80$ is the effective refractive index of the IBPF and $\theta_i$ is the angle of incidence. The corresponding relationship between $\theta_i$ and the transmitted wavelength $\lambda_c$ is shown in Fig.\ref{fig3}d. In our experiment, the IBPF was rotated in 1$^\circ$ increments from 0 to 42$^\circ$, yielding a maximum spectral step of $\Delta\lambda=2.3$nm, beyond which the transmitted signal was negligible.

The focal spot is imaged onto a high resolution CMOS sensor using a 20x long-working-distance objective and a 1 m focal length lens (Fig.\ref{fig3},$f_1$), yielding a magnification of approximately 100x. To retrieve the spatial phase for each spectral slice, a pick-off beamsplitter (Fig.\ref{fig3},BS) is inserted upstream of the light spring generation optics. The reference beam is apodized using an iris to a diameter less than 0.5 mm and injected off-axis into the focal spot imaging system at a lateral distance $d$ from the main beam path. Temporal overlap in the focal plane of lens $f_1$ is achieved by adjusting the optical delay line (DS1) in the reference arm. 

Each image acquired for a given wavelength contains the spatial interference pattern between the reference and light spring fields, as illustrated in Fig.\ref{fig3}e. The reference electric field at each wavelength, $E_{ref}(\lambda)$, is shaped by the near-field pinhole and thus has a spatial envelope significantly broader than that of the light spring field $E_{ff}(x,y,\lambda)$. As a result, the reference field can be approximated as spatially uniform with a flat phase profile. The recorded interference intensity is given by,
\begin{equation}
    I_{meas}(x,y,\lambda) \propto \left|E_{ff}\right|^2+\left|E_{ref}\right|^2+E^*_{ff}E_{ref}+E^*_{ref}E_{ff}
\end{equation}
To isolate the spatial phase of the light spring field, a 2D Fourier transform is applied to each interferogram. One of the off-axis (AC) terms is selected via masking in the Fourier domain and shifted to the origin before applying an inverse Fourier transform. The phase $\phi(x,y,\lambda)$ is then extracted using a 2D phase unwrapping algorithm, while the amplitude is computed as,
\begin{equation}
    \frac{|E^*_{ref}(\lambda)E_{ff}(x,y,\lambda) |^2}{|E_{ref}(\lambda)|^2}\approx|E_{ff}(x,y,\lambda) |^2\propto I_{ff}(x,y,\lambda)
\end{equation}
The resulting complex field at each wavelength is reconstructed as, 
\begin{equation}
    \psi_{ff}(x,y,\lambda)=\sqrt{\frac{I_{ff}(x,y,\lambda)}{\iint I_{ff}(x,y,\lambda)dA}}\exp{[i\phi(x,y,\lambda)]}.
\end{equation}

The complex field at each wavelength is then combined with the spectral phase $\varphi(\omega)$ and spectral amplitude $E_0(\omega)$ measured via FROG to reconstruct the full electric field in frequency space, as in Eq.\ref{eq1}. An inverse Fourier transform is then performed along each frequency at each transverse spatial point to obtain the electric field as a function of space and time, $E_{LS}(x,y,t)$. While it is possible to extend this reconstruction to include the propagation direction $z$ using Fourier optics propagators, such analysis is beyond the scope of this work.

\subsection*{Modal decomposition}
To compute the spectral-topological correlations, the measured far-field complex field must be decomposed into Laguerre-Gaussian (LG) modes. Each complex spatial field is first normalized such that $\langle\psi_{ff}(\lambda)|\psi_{ff}(\lambda)\rangle=1$, enabling the spectral-topological correlation to be computed as \cite{Longman:17}, 
\begin{equation}
    \mathcal{C}(\lambda,\ell)=\sum_p\left|\langle\psi_p^\ell|\psi_{ff}(\lambda)\rangle\right|^2
\end{equation}
where the LG modes in the focal plane are defined,
\begin{equation}
    \psi_p^\ell=\sqrt{\frac{2p!}{\pi(|\ell|+p)!}}\frac{1}{w_0}\left(\frac{\sqrt{2}r}{w_0}\right)^{|\ell|}\exp{\left(-\frac{r^2}{w_0^2}+i\ell\theta\right)}L_p^{|\ell|}\left(\frac{2r^2}{w_0^2}\right)
\end{equation}
with $w_0\approx f_{\#}\lambda$ representing the beam waist in the focal plane, and $L_p^{|\ell|}(X)$ the associated Laguerre polynomial. While a non-optimal choice of $w_0$ can increase the number of contributing radial modes $p$, it does not affect the decomposition of the azimuthal index $\ell$. 

However, the choice of the coordinate origin-specifically, where the radial coordinate $r$ is centered in the image-can introduce significant discrepancies in the calculated value of $\ell$. To illustrate this, consider the angular momentum density of a paraxial beam measured off-center from its propagation axis, 
\begin{equation}
    M_z=\epsilon_0\left[(\mathbf{r-r_0})\times\mathbf{E}\times\mathbf{B}\right]_z.
\end{equation}
This expression reveals two distinct contributions to the OAM: an intrinsic part and an extrinsic part. The intrinsic OAM, which arises from the internal structure of the field, is given by, 
\begin{equation}
    \ell_{\mathrm{int}}(\lambda)=-i\frac{\langle\psi_{ff}(\lambda)|\partial_\theta|\psi_{ff}(\lambda)\rangle}{\langle\psi_{ff}(\lambda)|\psi_{ff}(\lambda)\rangle}
\end{equation}
while the extrinsic OAM, resulting from the transverse displacement of the beam relative to the coordinate origin ($x_0,y_0$), is given by
\begin{equation}
    \ell_{\mathrm{ext}}(\lambda) = -i \frac{x_0 \langle \psi_{ff}(\lambda) | \partial_y | \psi_{ff}(\lambda) \rangle - y_0 \langle \psi_{ff}(\lambda) | \partial_x | \psi_{ff}(\lambda) \rangle}{\langle \psi_{ff}(\lambda) | \psi_{ff}(\lambda) \rangle}
\end{equation}
Incorrect selection of $x_0$ and $y_0$ can introduce spurious contributions to the measured spectral-topological correlations due to extrinsic OAM. This issue particularly pronounced in mixed OAM states, where the optical axis is not trivial to identify. To mitigate this, we determined the beam center using spectral regions dominated by a single, well-defined mode - typically found at the spectral extremities - and adopted this center as a common reference for all wavelength slices.  

Finally, we define the energy-normalized spectral-topological correlations as in units of mJ/$/\lambda$, 
\begin{equation}
    \tilde{\mathcal{C}}(\lambda,\ell)=\frac{\mathcal{C(\lambda,\ell)}}{\iint\mathcal{C}(\lambda,\ell)d\lambda d\ell}U_{0}
\end{equation}
where $U_0$ is the laser energy in the focus.

\subsection*{Off-axis spiral phase mirrors and dichroic beamsplitters}
The off-axis spiral phase mirrors were fabricated from fused silica substrates, with the helical surface profile directly polished into the optic using magneto-rheological finishing (MRF). An example surface profile map is given in Fig.\ref{fig3}a, revealing an RMS deviation of $\sim$5 nm from the designed phase structure. To accommodate the broad laser spectrum, the mirrors were coated with a high-bandwidth multi-layer dielectric coating centered at 800 nm, achieving a reflectivity exceeding $99.5\%$ across the entire spectral bandwidth.

\begin{figure}[h]
\centering
\includegraphics[width=\textwidth/2]{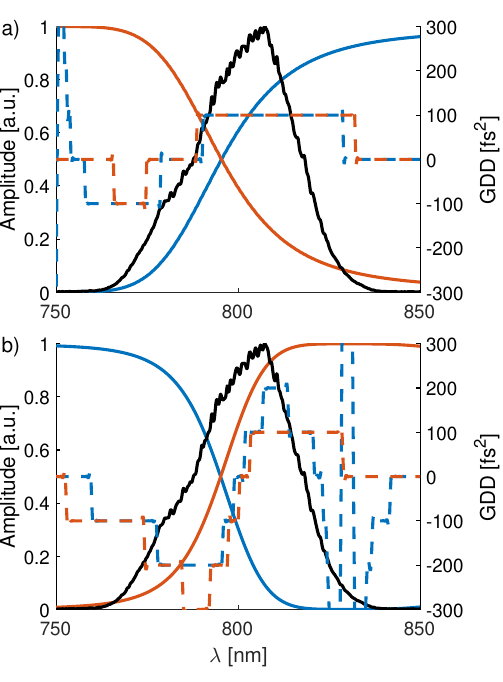}
\caption{Transmission, reflection, and group delay dispersion (GDD) curves for the dichroic beamsplitters used in the experiment. (a) Short-pass filter curves, with a slope parameter (see text) $\Lambda \approx \pm10.3$ nm and central wavelength $\lambda_0 = 797$ nm. (b) Long-pass filter with $\Lambda \approx \pm6.2$ nm and $\lambda_0 = 795$ nm. In both plots, the laser spectrum is shown in black, reflection in blue, and transmission in orange. Dashed lines indicate the corresponding GDD, referenced to the right axis.}
\label{fig4}
\end{figure}

The dichroic beamsplitters were fabricated by Alluxa, with their transmission, reflection, and GDD curves shown in Fig.\ref{fig4}a for the short pass-filter, and Fig.\ref{fig4}b for the long-pass data. In both plots, the measured laser spectrum is shown in black, while the reflection and transmission spectra are shown in blue and orange, respectively. Dashed lines indicate the GDD for each optic, referenced to the right vertical axis. 

The transmission and reflection characteristics of the filters can be modeled by fitting the data with a logistic function of the form: 
\begin{equation}
    T(\lambda),R(\lambda)=\left[1+\exp{\left(\frac{\lambda-\lambda_b}{\Lambda}\right)}\right]^{-1}
\end{equation}
where $\lambda_b$ is the central design wavelength, and $\Lambda$ controls the steepness and direction of the transition. For the short-pass filter (Fig.\ref{fig4}a), we find $\Lambda\approx\pm10.26$ nm and $\lambda_b=797$ nm. For the long-pass filter (Fig.\ref{fig4}b), $\Lambda\approx\pm6.2$ nm and $\lambda_b=795$ nm.

Using these slope parameters enabled a steep transition between reflection and transmission across the laser spectrum, while also minimizing the GDD. For the short-pass filter, the GDD remained at or below 100 fs$^2$ across all wavelengths of interest. For the long-pass filter, the GDD stayed at or below 200 fs$^2$ for the majority of the spectral bandwidth.

\bibliography{references}

\section*{Acknowledgements}

This work was performed under the auspices of the U.S. Department of Energy by Lawrence Livermore National Laboratory under Contract No. DE-AC52-07NA27344,  This work was supported in part by the Laboratory Directed Research and Development program under 24-ERD-031 and 25-LW-113, the National Science Foundation under Grants No. DMR-1548924 and PHY-1753165, based upon work supported by the U.S. Department of Energy under Grant No. DE-SC0023504, as well as supported by the Natural Sciences and Engineering Research Council of Canada (Grant No. RGPIN-2019-05013). This material is The authors would like to thank P. Michel, E. Kur, N. Lemos, and J. M. Di-Nicola for helpful discussions. 

\section*{Author contributions statement}
A.L. and R.F. conceived the concept experiments. A.L. and E.G. conceived the measurment method.  A.L, D.A, E.G, C.G, and F.D, conducted the experiments. A.L, T.S, G.T. and C.H. designed and fabricated the spiral phase optics. A.L, E.G. and D.A. analyzed the results. A.L. wrote the manuscript. All authors reviewed the manuscript.

\section*{Additional information}
The authors declare no competing financial interests.
\end{document}